\documentclass[twocolumn,showpacs,preprintnumbers,amsmath,amssymb,floatfix,nofootinbib]{revtex4}

\usepackage{verbatim}
\usepackage{enumerate}
\usepackage{graphicx}
\usepackage{dcolumn}
\usepackage{bm}
\usepackage{color}
\usepackage{url}
\usepackage{amsmath}
\usepackage{morefloats}
\usepackage[normalem]{ulem}

\usepackage{color}

\graphicspath{{./figures/}}
\newcommand{\be}{\begin{equation}}
\newcommand{\ee}{\end{equation}}
\newcommand{\ba}{\begin{eqnarray}}
\newcommand{\ea}{\end{eqnarray}}
\newcommand{\nn}{\nonumber}

\newcommand{\hyp}{\mathcal{H}}
\def\ltsima{$\; \buildrel < \over \sim \;$}
\def\simlt{\lower.5ex\hbox{\ltsima}}
\def\gtsima{$\; \buildrel > \over \sim \;$}
\def\simgt{\lower.5ex\hbox{\gtsima}}
\begin{document}

\title[blah]{TIGER: A data analysis pipeline for testing the strong-field dynamics of general relativity with gravitational wave signals from coalescing compact binaries}

\author{M.~Agathos$^{1}$} 
\email{magathos@nikhef.nl}
\author{W.~Del Pozzo$^{1,2}$}
\author{T.G.F.~Li$^{1,3}$}
\author{C.~Van Den Broeck$^{1}$}
\author{J.~Veitch$^{1}$}
\author{S.~Vitale$^{4}$}
\affiliation{$^1$Nikhef -- National Institute for Subatomic Physics, Science Park 105, 1098 XG Amsterdam, The Netherlands \\
$^2$School of Physics and Astronomy, University of Birmingham, Edgbaston, Birmingham B15 2TT, United Kingdom\\
$^3$LIGO Laboratory, California Institute of Technology, Pasadena, CA 91125, USA\\
$^4$LIGO Laboratory, Massachusetts Institute of Technology, Cambridge, MA 02139, USA}
\date{\today}

\begin{abstract}
The direct detection of gravitational waves with upcoming second-generation gravitational wave observatories such as Advanced LIGO and Advanced Virgo will allow us to probe the genuinely strong-field dynamics of general relativity (GR) for the first time. We have developed a data analysis pipeline called TIGER (Test Infrastructure for GEneral Relativity), which uses signals from compact binary coalescences to perform a model-independent test of GR. 
In this paper we focus on signals from coalescing binary neutron stars, for which sufficiently accurate waveform models are already available which can be generated fast enough on a computer that they can be used in Bayesian inference. By performing numerical experiments in stationary, Gaussian noise, we show that for such systems, TIGER is robust against a number of unmodeled fundamental, astrophysical, and instrumental effects, such as differences between waveform approximants, a limited number of post-Newtonian phase contributions being known, the effects of neutron star tidal deformability on the orbital motion, neutron star spins, and instrumental calibration errors. 
\end{abstract}


\maketitle

\section{Introduction and overview}

General relativity (GR) is a highly non-linear, dynamical theory of gravity. Yet, until the 1970s, almost all of its tests were based on the behavior of test particles in a \emph{static} gravitational field \cite{MTW}, such as the perihelion precession of Mercury, the deflection of starlight by the Sun, and Shapiro time delay. The parameterized post-Newtonian (PPN) formalism (for an overview, see \cite{Will2006}) was developed as a systematic framework for these and other tests; even so, the interpretation of most of the available data did not require much more than an expansion of the Schwarzschild metric in $GM/(c^2 r)$, with $M$ the mass and $r$ the distance, up to the first few non-trivial orders. Although excellent agreement with theory was obtained, the tests that were actually performed amounted to little more than probing the effect on the motion of test masses of low-order general relativistic corrections to the Newtonian gravitational field.

The situation improved with the discovery of the Hulse-Taylor binary neutron star in 1974 \cite{ht75}. One of the components could be observed electromagnetically as a pulsar, and this way it was inferred that the binary loses energy and angular momentum through gravitational wave (GW) emission as predicted by GR, at least at the level of the quadrupole formula \cite{Weisberg2010}. Subsequently, more relativistic binaries were discovered, allowing for impressive new tests of GR in a parameterized post-Keplerian (PPK) framework \cite{Freire2012}. However, if one is interested in further probing the dissipative dynamics of binaries, and especially the dynamics of spacetime itself, what matters is the \emph{orbital compactness} $GM/(c^2 R)$ (with $M$ the total mass and $R$ the separation), as well as the orbital velocity $v/c$. Even the newly discovered neutron star-white dwarf system \cite{Antoniadis2013} only has $GM/(c^2 R) \sim 2 \times 10^{-6}$, and $v/c \sim 4 \times 10^{-3}$. For comparison, the surface gravity of the Sun is $G M_\odot/(c^2 R_\odot) \sim 10^{-6}$, and the orbital velocity of Mercury is $v/c \sim 1.6 \times 10^{-4}$. 

By contrast, binaries consisting of neutron stars and/or black holes on the verge of merger will have $G M/(c^2 R) > 0.2$ and $v/c > 0.4$, with copious gravitational wave emission. Being able to observe the orbital motion of such systems would give us access to the genuinely strong-field, relativistic regime of gravity. Most importantly, we would like to probe the dynamical self-interaction of spacetime itself, such as the scattering of quadrupolar waves off the Schwarzschild curvature generated by the binary as a whole \cite{Blanchet1994,Blanchet1995}. The only way to gain empirical access to such phenomena is through direct gravitational wave detection. 

A network of second-generation gravitational wave detectors is currently under construction. The Advanced LIGO \cite{aLIGO} and Advanced Virgo \cite{AdV}  GW observatories are expected to start taking data in 2015, with gradual upgrades in the following years. The smaller GEO-HF in Germany is already active \cite{GEO}. KAGRA \cite{KAGRA} in Japan and possibly LIGO-India \cite{IndIGO} will come online a few years later. These detectors may find tens of GW signals per year from coalescing compact binaries composed of two neutron stars (BNS), a neutron star and a black hole (NSBH), or two black holes (BBH). The predicted detection rates for the Advanced LIGO-Virgo network are in the range $1 - 100\,\mbox{yr}^{-1}$ depending on the astrophysical event rate, the instruments' duty cycle, and the sensitivity evolution of the detectors \cite{rates,commissioning}; see also \cite{Chen2012} for detection rate predictions assuming that short, hard gamma ray bursts are caused by coalescing binaries. 

There is a considerable body of literature on the constraints that can be put on various \emph{specific} alternative theories of gravity with ground-based and space-based GW detectors, and pulsar timing arrays; see \cite{Gair2012,Yunes2013} and references therein. What we will be interested in here are \emph{model-independent} tests of GR itself. A first step in that direction was taken by Arun \emph{et al.}~\cite{Arun2006a,Arun2006b,Mishra2010} in the context of compact binary inspiral. Their method exploits the fact that, at least for binaries where neither component has spin, all coefficients $\psi_i$ in the post-Newtonian (PN) expansion of the inspiral phase (see below for their definition) only depend on the component masses $m_1$, $m_2$. Hence only two of them are independent, and a comparison of any three of them allows for a test of GR. Such a method would be extremely general, in that one does not have to look for any specific way in which GR might be violated; instead, very generic deviations can 
be searched for. A similar idea was pursued in the context of ringdown by Gossan \emph{et al.}~\cite{Gossan2012}: if the No Hair Theorem applies to Nature, then the frequencies $f_{nlm}$ and damping times $\tau_{nlm}$ of the various ringdown modes again only depend on two quantities, in this case the mass $M$ and spin $J$ of the final black hole. 

The original ideas of \cite{Arun2006a,Arun2006b,Mishra2010} have the drawback that they rely on parameter estimation, which makes it difficult to combine information from multiple sources. An alternative way of testing GR is \emph{Bayesian model selection}. Here one compares two hypotheses, one corresponding to the GW waveform model predicted by GR, and the other to a model which has parameterized deformations of the GR waveform, characterized by additional parameters $\{\delta\chi_1, \delta\chi_2, \ldots, \delta\chi_{N_T}\}$. This was the approach taken by Del Pozzo \emph{et al.}~\cite{DelPozzo2011} in the context of inspiral (where a single additional parameter was introduced, related to the graviton mass),  and again by Gossan \emph{et al.} for ringdown (where multiple extra free parameters were considered) \cite{Gossan2012}. Yunes and collaborators \cite{Yunes2009,Cornish2011,Chatziioannou2012} proposed a parameterization of non-GR waveforms guided by the ways in which a variety of alternative theories of gravity modify the GR 
waveform, leading to the ``parameterized post-Einsteinian" (PPE) framework. For the relationship between the PPN, PPK, and PPE formalisms, see \cite{Sampson2013}.

In the abovementioned Bayesian studies, a comparison was made between a waveform model in which all the extra parameters $\delta\chi_i$ were allowed to vary, and a waveform model where all of them took their GR values (which for the present discussion we can take to mean $\delta\chi_i = 0$ for $i = 1, \ldots, N_T$). As noted by Li \emph{et al.}~\cite{Li2012a}, this corresponds to asking the question ``Do \emph{all} of the $\delta\chi_i$ differ from zero at the same time?" Let us denote the associated hypothesis by $H_{1 2 \ldots N_T}$, which is to be compared with the GR hypothesis $\hyp_{\rm GR}$.  A more general (and hence more interesting) question is: ``Do \emph{one or more} of the $\delta\chi_i$ differ from zero?" Denote the corresponding hypothesis by $\hyp_{\rm modGR}$. As shown in \cite{Li2012a}, although there is no single waveform model associated with $\hyp_{\rm modGR}$, testing the latter amounts to testing $2^{N_T} - 1$ disjoint sub-hypotheses $H_{i_1 i_2 \ldots i_k}$ corresponding to all subsets $\{\delta\chi_{i_1}, \delta\chi_{i_2}, \ldots, \delta\chi_{i_k}\}$ of the full set of ``testing parameters" $\{\delta\chi_1, \delta\chi_2, \dots, \delta\chi_{N_T}\}$. A given $H_{i_1 i_2 \ldots i_k}$ is tested by a waveform model in which $\delta\chi_{i_1}, \delta\chi_{i_2}, \ldots, \delta\chi_{i_k}$ are free, but all the other $\delta\chi_j$ are fixed to zero. The Bayes factors against GR for all of these sub-hypotheses can be combined into a single \emph{odds ratio} which compares $\hyp_{\rm modGR}$ with $\hyp_{\rm GR}$. 

In the present paper, we will consider deformations in the inspiral phase of the waveform, which in the stationary phase approximation \cite{Thorne1987,Sathyaprakash1991} takes the form 
\be
\Psi(f) = 2\pi f t_c - \varphi_c - \frac{\pi}{4} + \sum_{j=0}^7 \left[ \psi_j  + \psi_j^{(l)}\ln f \right]\,f^{(j - 5)/3},
\label{phase}
\ee
where $t_c$ and $\varphi_c$ are, respectively, the time and phase at coalescence, and in GR, the coefficients $\psi_j$, $\psi_j^{(l)}$ are specific, known functions of the component masses $m_1$, $m_2$ and spins $\vec{S}_1$, $\vec{S}_2$. Parameterized deformations of the phase can be introduced by writing $\psi_i = [1+ \delta\chi_i]\psi_i^{\rm GR}$, where $\psi_i^{\rm GR} = \psi_i^{\rm GR}(m_1, m_2, \vec{S}_1, \vec{S}_2)$ is the expression for $\psi_i$ as a function of component masses $m_1$, $m_2$ and spins $\vec{S}_1$, $\vec{S}_2$ that GR predicts.\footnote{One could also introduce parameterized deformations of the amplitude, but studies have shown that for low-mass systems, second-generation detectors will not be very sensitive to sub-leading effects in the amplitude \cite{VanDenBroeck2007a,VanDenBroeck2007b,OShaughnessy2013}.}

Given a catalog of sources $d_1, d_2, \ldots d_\mathcal{N}$ detected with the dedicated search pipelines \cite{Beauville2005,Brown2005,Babak2006,Cokelaer2007a,Cokelaer2007b,VanDenBroeck2009,Brown2012,Babak2013,Brown2013}, assuming equal prior odds for all the sub-hypotheses $H_{i_1 i_2 \ldots i_k}$ and taking the data streams for the individual detections to be independent, the odds ratio for $\hyp_{\rm modGR}$ against $\hyp_{\rm GR}$ yields \cite{Li2012a,Li2012b,VanDenBroeck2013}
\ba
\mathcal{O}^{\rm modGR}_{\rm GR} &\equiv& \frac{P(\hyp_{\rm modGR}|d_1,\ldots, d_\mathcal{N}, I)}{P(\hyp_{\rm GR}|d_1, \ldots, d_\mathcal{N}, I)} \nn\\
&=& \frac{\alpha}{2^{N_T}-1} \sum_{i_1 < \ldots < i_k; k \leq N_T} \prod_{A = 1}^\mathcal{N} \frac{P(d_A|H_{i_1 \ldots i_k}, I)}{P(d_A|\hyp_{\rm GR}, I)},\nn\\
\label{logO}
\ea
with $P(d_A|H_{i_1 \ldots i_k}, I)$ and $P(d_A|\hyp_{\rm GR}, I)$ the evidences for $H_{i_1 i_2 \ldots i_k}$ and $\hyp_{\rm GR}$, respectively; $I$ denotes any background information we may hold, and $\alpha = P(\hyp_{\rm modGR}|I)/P(\hyp_{\rm GR}|I)$ is the ratio of prior odds for $\hyp_{\rm modGR}$ against $\hyp_{\rm GR}$.

If GR happens to be valid then one would expect $\mathcal{O}^{\rm modGR}_{\rm GR} < 1$, or $\ln \mathcal{O}^{\rm modGR}_{\rm GR} < 0$. However, the noise in the detectors can mimic violations of GR, so that one can have $\ln \mathcal{O}^{\rm modGR}_{\rm GR} > 0$ even if GR is in fact the correct theory of gravity. Moreover, there will be some effect of numerical inaccuracy in the calculation of the log odds ratio. To make sure that we will not erroneously declare a GR violation, the measured log odds ratio will be compared with a \emph{background distribution}. The latter is constructed by taking a large number of simulated GR signals, all having different masses, sky locations, orientations, and distances picked from astrophysically motivated distributions (see Sec.~\ref{sec:setup} below), and injecting them into stretches of data surrounding the ones the detections are in, to have similar noise realizations. Here one can adopt the treatment of ``on-source" and ``off-source" data as in searches for gravitational wave events associated with gamma ray bursts; see \cite{S6GRB} and references therein. These injections can be combined randomly into ``catalogs", each containing however many sources were observed in reality. For every catalog of background injections one can calculate $\ln\mathcal{O}^{\rm modGR}_{\rm GR}$, arriving at an estimate for the  distribution of the log odds ratio for the case where GR is correct. Given such a distribution and picking a maximum tolerable false alarm probability, a threshold can be computed for the \emph{measured} log odds ratio to overcome.

For details of the above definitions and derivations, we refer to \cite{Li2012a,Li2012b,VanDenBroeck2013}. As explained in those references and further elucidated in this paper, the approach of Li \emph{et al.}~has several attractive features: 
\begin{enumerate}[(i)]
\item One can use an arbitrarily large number of ``testing parameters" without having to worry about a model being insufficiently parsimonious in cases where the true number of non-GR parameters is small, due to the availability of sub-hypotheses corresponding to different numbers of free parameters. 
\item Information from multiple sources can trivially be combined, leading to a stronger test of GR. 
\item It is well-suited to a regime where most sources have a small signal-to-noise ratio, again because of the use of multiple non-GR sub-hypotheses.
\item  It will allow us to find a wide range of deviations from GR, even ones that are well outside the particular parameterized waveform family used.
\item The method is not tied to any given waveform model, or even any particular part of the coalescence process.  
\end{enumerate}

Given these advantages, it is natural to take the above scheme as a basis for computer code to test GR using actual detector data. Such a data analysis pipeline is now in place within the LIGO Algorithm Library \cite{LAL}. It is called \emph{TIGER}, for ``Test Infrastructure for GEneral Relativity". 

Before we can be sure of the usefulness of TIGER in a realistic data analysis setting, we must check its robustness against any unknown fundamental, astrophysical, and instrumental effects. We focus on BNS, since for this case, waveform models that accurately capture the relevant physics \emph{and} can be generated sufficiently fast on a computer have been available for some time now \cite{Buonanno2009}. In practice, BNS systems could be selected for by looking at the \emph{chirp mass} $\mathcal{M} = M \eta^{3/5}$, a parameter which tends to be very well determined in gravitational wave parameter estimation, with uncertainties of a few percent.\footnote{Here we are referring to both statistical and systematic errors; see \cite{S6PE} for examples.} In Dominik \emph{et al.} \cite{Dominik2012}, results from a large number of formation models for compact binaries are given. They find the minimum chirp mass for NSBH to be $1.7\,M_\odot$, and $2.4\,M_\odot$ for BBH. Thus, selecting only detections for which \emph{e.g.}~$\mathcal{M} < 1.3\,M_\odot$ at 95\% confidence should remove all NSBH and BBH events.
Of course, it is entirely possible that some genuine BNS detections will be removed in this way (in fact, this is what the BNS results of \cite{Dominik2012} suggest), but the procedure is a conservative one. We note that of necessity, the selection will have to be done based on parameter estimation with \emph{GR waveforms}. If GR is incorrect, then there could be a large bias in the measurement of (among other parameters) $\mathcal{M}$  \cite{Yunes2009,DelPozzo2011,Li2012a,Vitale2014}, in which case even a BBH system could be mis-classified as a BNS system. However, in that case we expect TIGER to \emph{a fortiori} indicate a violation of GR. 


Focusing on BNS, the following issues need to be addressed:
\begin{enumerate}[(i)]
\item Even for binary neutron star coalescence, there are small differences between the various waveform approximants that are available. Since TIGER is specifically designed to find anomalies in the signals, we must make sure that these discrepancies, however minor, are not mistaken for violations of GR.
\item Post-Newtonian waveforms are only available up to 3.5PN in phase. What might be the effect of unknown PN contributions?
\item In the final stages of inspiral, neutron stars get deformed because of each other's tidal fields. This has an effect on the orbital motion, which gets imprinted onto the GW signal waveform. The size of these tidal effects is set by the neutron star equation of state, about which currently not much is known. Can we avoid mistaking unknown tidal effects for a violation of GR?
\item The dimensionless spins of neutron stars in binaries are generally expected to be quite small, but the resulting spin-orbit and spin-spin effects will nevertheless need to be taken into account. 
\item The calibration of the instruments will be imperfect, leading to frequency dependent uncertainties in the interpretation of amplitudes and phases. What will their impact be?
\end{enumerate}

In order to see how these effects can be brought under control, we perform numerical experiments in simulated stationary, Gaussian noise following the predicted noise curves of Advanced LIGO and Advanced Virgo at their final design sensitivities \cite{aLIGO,AdV}. Note that in reality, the noise will be neither Gaussian nor stationary due to ``glitches". As explained above, TIGER involves the calculation of a background distribution in which these additional unknowns will be included automatically, possibly resulting in a widening of the background. However, here we focus on the points above; further instrumental issues will be dealt with in a forthcoming study.

This paper is structured as follows. In Sec.~\ref{sec:setup}, we explain the setup of the simulations, and how we will compare results arising from different assumptions. The main results are presented in Sec.~\ref{sec:robustness}, where we show how TIGER can be made robust against differences between waveform approximants, limited availability of post-Newtonian phase contributions, unknown neutron star tidal deformability, instrumental calibration errors, and the effects of neutron star spins.  Conclusions and future directions are discussed in Sec.~\ref{sec:discussion}.

Unless stated otherwise, we will use units such that $G = c = 1$.

\section{Setup of the simulations and comparison of results}
\label{sec:setup}


The results in this paper pertain to simulations of BNS signals in stationary, Gaussian noise following the design sensitivity of Advanced LIGO and Advanced Virgo \cite{aLIGO,AdV}. Component masses were in the range $1 -2\,M_\odot$,\footnote{This means that in the simulations, we don't quite use the cut $\mathcal{M} < 1.3\,M_\odot$ proposed above, but our key results are unlikely to be affected by this choice.} sky positions and orientations were chosen from uniform distributions on the sphere, and sources were placed uniformly in co-moving volume with luminosity distance $D \in [100, 250]$ Mpc. Depending on the type of robustness test, the signal waveform was taken to be TaylorF2 with zero or (anti-)aligned spins, or TaylorT4 with precessing spins; the recovery was done with TaylorF2 waveforms, again with either zero or (anti-)aligned spins. Only sources with optimal network SNR above 8 were taken into account \cite{Cutler1993}. Occasionally it would happen that a source survived the SNR cut without being found by the GR 
waveform model, meaning $\ln B^{\rm GR}_{\rm noise} \simeq 0$, with $B^{\rm GR}_{\rm noise} \equiv P(d|\hyp_{\rm GR}, I)/P(d|\hyp_{\rm noise}, I)$ the log Bayes factor for the hypothesis of a GR signal being present against the noise-only hypothesis. Such sources were discarded by imposing $\ln B^{\rm GR}_{\rm noise} > 32$, motivated by the fact that the main contribution to $\ln B^{\rm GR}_{\rm noise}$ is $(1/2)\,\langle h_{\rm GR}|h_{\rm GR}\rangle = (1/2)\,\mbox{SNR}^2$, with $h_{\rm GR}$ the GR waveform, and $\langle \,\cdot\,|\,\cdot\, \rangle$ is the usual noise-weighted inner product \cite{Maggiore}:  
\be
\langle a | b \rangle \equiv 4 \Re \int_{f_0}^{f_{\rm LSO}} df \,\frac{\tilde{a}^\ast(f)\,\tilde{b}(f)}{S_n(f)},
\ee
where a tilde denotes the Fourier transform, and $S_n(f)$ is the one-sided noise power spectral density. To compute the evidences $P(d|H_{i_1 i_2 \ldots i_k}, I)$ and $P(d|\hyp_{\rm GR}, I)$ we used the nested sampling method as implemented by Veitch and Vecchio \cite{Veitch2008a,Veitch2008b,Veitch2010}, with 1000 ``live points" and 100 ``MCMC points", which leads to an uncertainty $\lesssim 1$ in log Bayes factors against noise \cite{Veitch2010}.

In what follows, we will want to compare different background distributions: with or without calibration errors, with or without tidal effects in the injections, and so on. A convenient way of quantifying the difference between distributions is by means of the Kolmogorov-Smirnov statistic \cite{Kolmogorov1933,Smirnov1948}. Consider backgrounds $P(\ln\mathcal{O} | \hyp_{\rm GR}, \kappa_1, I)$ and $P(\ln\mathcal{O} | \hyp_{\rm GR}, \kappa_2, I)$ for different injection sets $\kappa_1$, $\kappa_2$ (or in the case of calibration errors, different simulated data sets containing injections). Construct the \emph{cumulative} distributions of log odds ratio and call these $F_{1, N}(\ln\mathcal{O})$ and $F_{2, N'}(\ln\mathcal{O})$, respectively; here $N$ and $N'$ are the numbers of log odds ratio values that are available in each of the two cases. Then the Kolmogorov-Smirnov (KS) statistic is just the largest distance between the cumulative distributions:
\be
D_{N,N'}^{1,2} \equiv \mbox{sup}_{\ln\mathcal{O}} |F_{1, N}(\ln\mathcal{O}) - F_{2, N'}(\ln\mathcal{O})|.
\label{KS}
\ee
Note that by construction, this is a number between 0 and 1. If $D_{N,N'}^{1,2} \ll 1$, then the difference between the background distributions can be considered small.

\section{Robustness of TIGER against unknown fundamental, astrophysical, and instrumental effects}
\label{sec:robustness}

We now show how the TIGER pipeline can be made robust against effects of a fundamental, astrophysical, or instrumental nature which can not easily be accounted for in our waveform models. In turn, we study the impact of neutron star tidal deformability,  differences between waveform approximants, unknown contributions to the phase at high PN order, instrumental calibration errors, the effect on the background of the number of coefficients used, and precessing neutron star spins. We expressly gauge the importance of each of these issues separately, in order to clearly demonstrate how each of them can be brought under control, before finally considering the situation where all of them are jointly present. 

\subsection{Neutron star tidal deformability}
\label{subsec:tidal}

As two neutron stars spiral towards each other, each will get deformed due to the tidal field of the other. These deformations have an influence on the orbital motion which gets imprinted onto the emitted gravitational wave signal. The size of the effect is set by the \emph{tidal deformability} $\lambda({\rm EOS}, m)$, which relates the Newtonian tidal tensor $\mathcal{E}_{ij}$ of one star to the induced quadrupole moment $Q_{ij}$ of the other: $Q_{ij} = -\lambda({\rm EOS}, m)\,\mathcal{E}_{ij}$. One has $\lambda(m) = (2/3) k_2(m) R^5(m)$, with $k_2$ the second Love number and $R$ the neutron star radius. As the notation suggests, the tidal deformability depends on mass in a way that is determined by the neutron star equation of state (EOS). In the presence of tidal effects, the waveform phase takes the form $\Phi(v) = \Phi_{\rm PP}(v) + \Phi_{\rm tidal}(v)$, where $\Phi_{\rm PP}(v)$ is the usual point particle contribution, and to 1PN beyond leading order for tidal contributions one has\footnote{Damour, 
Nagar, and Villain have computed tidal contributions to 2.5PN \cite{Damour2012}, but these were not yet available when this work was started.} \cite{Hinderer2009}
\begin{widetext}
\begin{equation}
\Phi_{\rm tidal}(v) 
= \sum_{a=1}^2 \frac{3\lambda_a}{128 \eta M^5}\left[ -\frac{24}{\chi_a} \left(1 + \frac{11 \eta}{\chi_a}\right)\, v^5 -\frac{5}{28 \chi_a} \left(3179 - 919\,\chi_a - 2286\,\chi_a^2 + 260\,\chi_a^3 \right)\,v^7
\right].
\label{Phitidal}
\end{equation} 
\end{widetext}
The sum is over the components of the binary, and $\lambda_a = \lambda(m_a)$, $\chi_a = m_a/M$ for $a = 1, 2$. Note that although these contributions occur at 5PN and 6PN in the phase, they come with a prefactor that is potentially quite large: $\lambda/M^5 \propto (R/M)^5 \sim 10^2 - 10^5$ \cite{Lattimer2013}, so that the effect can be noticeable even with second-generation detectors. Indeed, in \cite{DelPozzo2013} it was shown that, if one assumes GR to be correct, the EOS can be significantly constrained by combining information from $\mathcal{O}(20)$ BNS observations. This in turn means that tidal effects could be mistaken for GR violations. 

Since little is known about the EOS -- in fact, currently the tidal deformability is uncertain by an order of magnitude -- we have no way of including an accurate description of it in our waveform models. However, because of the high PN order at which these effects occur, they will only be important at very high frequencies. Indeed, as shown by Hinderer \emph{et al.}~\cite{Hinderer2009} (see also the recent work by Read \emph{et al.} \cite{Read2013}), with second-generation detectors they only become noticeable for $f > 450$ Hz. For this reason we terminate our template waveforms at $f = 400$ Hz (which in terms of characteristic velocity and compactness corresponds to $v/c \sim 0.25$ and $GM/(c^2 R) \sim 0.07$, respectively). As it turns out, this leads to a loss in SNR of less than a percent, and in any case TIGER mostly probes the lower PN orders, corresponding to lower frequencies. However, here too we want to explicitly check that this suffices to make TIGER impervious to the unknown effect. 

\begin{figure}[!h]
\includegraphics[width=\columnwidth]{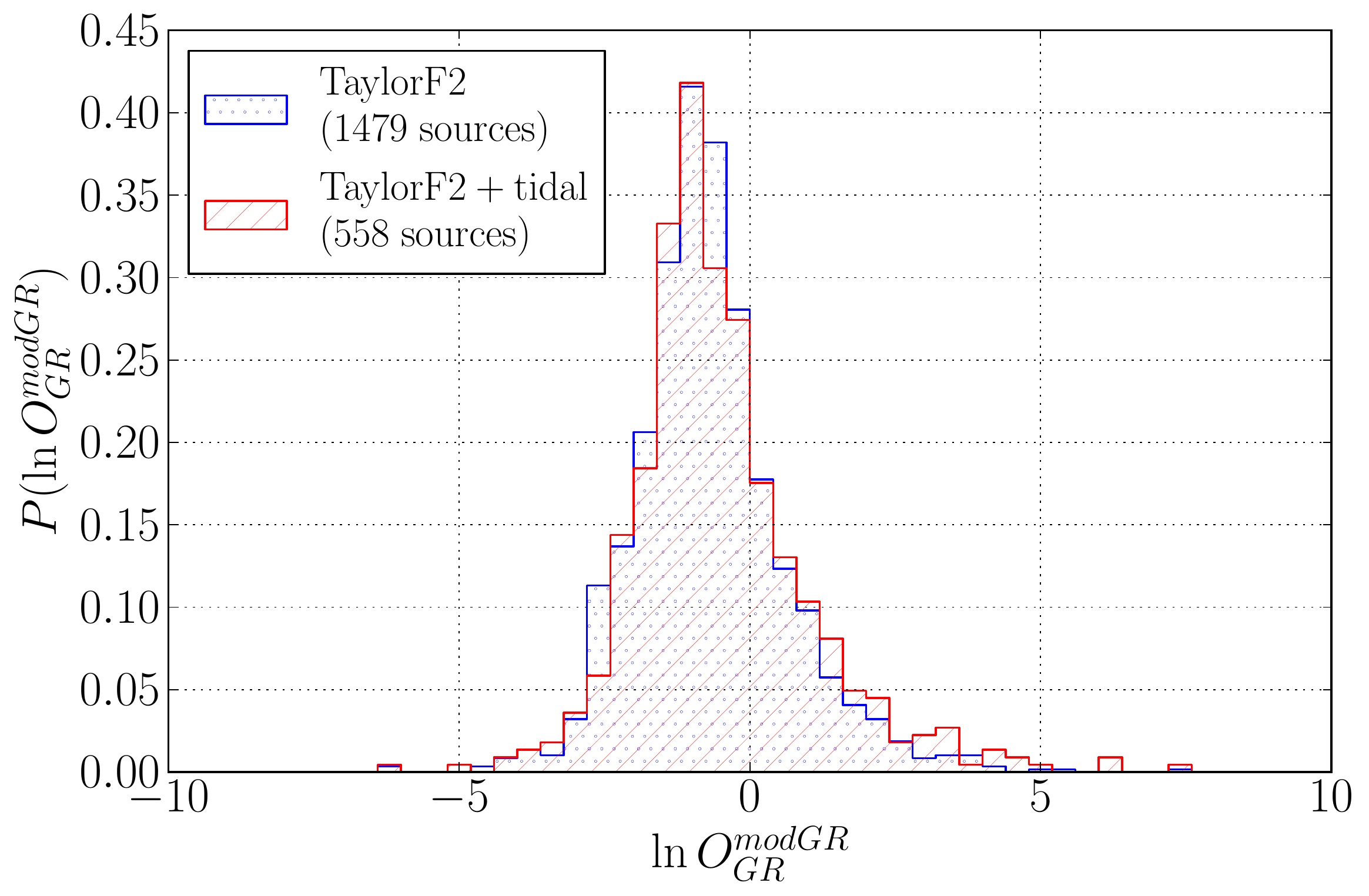}
\caption{Single-source background distributions for TaylorF2 injections without tidal effects (blue, dotted) and injections with strong tidal deformability (red, dashed), both analyzed with TaylorF2 waveforms that are cut off at $f = 400$ Hz.} 
\label{fig:tidal}
\end{figure}

In Fig.~\ref{fig:tidal}, we compare the background  for TaylorF2 injections without tidal effects, with the background obtained from injections with a very hard EOS (corresponding to large deformability), namely the one labeled MS1 in \cite{Hinderer2009}. The injected waveforms are taken to terminate at LSO while the recovery waveforms (also TaylorF2) are cut off at 400 Hz in both cases. 

Consider the background distribution for ``point particle" (PP) injections $\kappa_{\rm PP}$ (no tidal effects), $P(\ln\mathcal{O} | \hyp_{\rm GR}, \kappa_{\rm PP}, I)$, and the distribution of log odds ratio for MS1 injections $\kappa_{\rm MS1}$, $P(\ln\mathcal{O} | \hyp_{\rm GR}, \kappa_{\rm MS1}, I)$. Using the cumulative distributions of log odds in the two cases, one can construct the Kolmogorov-Smirnov statistic as in Eq.~(\ref{KS}). For the injection sets $\kappa_{\rm PP}$, $\kappa_{\rm MS1}$ used in Fig.~\ref{fig:tidal}, we find $D_{N,N'}^{\rm PP, MS1} = 0.06$, indicating that the two background distributions are very close to each other. We conclude that the 400 Hz cut-off renders tidal effects invisible without affecting TIGER's ability to look for GR violations.  

Below we will continue to implement a 400 Hz cut-off in the recovery waveforms.

\subsection{Differences between waveform approximants}
\label{subsec:waveforms}

For all the post-Newtonian waveform approximants, the phase $\phi(t)$ and instantaneous velocity $v(t)$ (or equivalently $t(v)$) are computed from the conserved energy per unit mass $E(v)$ and the gravitational wave flux $\mathcal{F}(v)$ through Kepler's law and the flux-energy balance equation:
\ba
\frac{d\phi}{dt} - \frac{v^3}{M} &=& 0, \label{Kepler}\\
\frac{dv}{dt} + \frac{\mathcal{F}}{M E'(v)} &=& 0, \label{energyflux}
\ea 
where the prime denotes derivation with respect to $v$. The solutions take the general form
\ba
t(v) &=& t_{\rm ref} + M \int_v^{v_{\rm ref}} dv\,\frac{E'(v)}{\mathcal{F}(v)}, \label{tv} \\
\phi(v) &=& \phi_{\rm ref} + \int_v^{v_{\rm ref}} dv\, v^3 \frac{E'(v)}{\mathcal{F}(v)}, \label{phiv}
\ea
where $t_{\rm ref}$ and $\phi_{\rm ref}$ are integration constants, and $v_{\rm ref}$ is an arbitrary reference velocity. Now, since $E(v)$ and $\mathcal{F}(v)$ are known as series expansions in $v$ up to a finite order, there are multiple ways of treating the above equations. In the case of the so-called TaylorT1 approximant, $E'(v)/\mathcal{F}(v)$ is kept as a ratio of polynomials, and Eqns.~(\ref{Kepler}) and (\ref{energyflux}) are solved numerically. In the case of TaylorT4, what one does instead is to expand the ratio $E'(v)/\mathcal{F}(v)$ and truncate the result at the consistent PN order, after which Eqns.~(\ref{Kepler}), (\ref{energyflux}) are again solved numerically. TaylorT2 is obtained by expanding and consistently truncating $E'(v)/\mathcal{F}(v)$,  and integrating Eqns.~(\ref{tv}) and (\ref{phiv}) to obtain a pair of transcendental equations for $\phi$ and $t$ as functions of $v$, which are then solved numerically. For TaylorT3 one also expands and truncates $E'(v)/\mathcal{F}(v)$, and 
integrates Eqns.~(\ref{tv}) and (\ref{phiv}) to obtain expressions for $\phi(v)$ and $t(v)$. The latter is inverted to $v(t)$, and a representation of $\phi(t) = \phi(v(t))$ is computed. Finally, the frequency domain TaylorF2 approximant is obtained through the \emph{stationary phase approximation}, by utilizing a saddle point in the calculation of the Fourier transform of the time domain waveform. For more details on all these approximants, see \cite{Buonanno2009} and references therein.

A qualitatively different way of obtaining waveform models is the effective-one-body (EOB) method. Here a mapping is established between the motion of the two component masses and the motion of a \emph{single} particle in an effective metric, which is captured by a set of Hamiltonian equations for the angular and radial motion. These are solved numerically. The advantage of this method is that the resulting waveforms are reliable up to later times compared to the PN ones (well into the plunge preceding merger), which also means that they lend themselves particularly well to being further ``tuned" using input from numerical simulations after being completed with a ringdown waveform. Here too we point to \cite{Buonanno2009} and references therein for further information. 

The authors of \cite{Buonanno2009} calculated the \emph{effectualness} and \emph{faithfulness} of post-Newtonian waveforms with respect to each other, as well as with an EOB waveform model tuned using numerical simulations, and this for a variety of component masses. The effectualness is a measure of how effective a waveform model $h_t$ will be when used as a template to detect a ``signal" waveform $h_s$; for given intrinsic and extrinsic signal parameters $\vec{\lambda}$, it is defined as $\max_{\vec{\theta}} \langle \hat{h}_s(\vec{\lambda}) | \hat{h}_t(\vec{\theta}) \rangle$, where $\hat{h} \equiv h/\sqrt{\langle h|h \rangle}$, and $\langle \,\cdot\,|\,\cdot\, \rangle$ is again the usual noise-weighted inner product \cite{Maggiore}. In the case of faithfulness, the intrinsic parameters $\vec{\lambda}_{\rm intr}$ of ``signal" and ``template" are taken to be the same, and the maximization is only over the template's time and phase at coalescence: $\max_{t_c, \varphi_c} \langle \hat{h}_s(\vec{\lambda}_{\rm intr})|\hat{h}_t(\vec{\lambda}_{\rm intr}) \rangle$. 
In the expected mass range of NSBH and BBH, there can be significant differences between the PN approximants amongst themselves, and with EOB waveforms. However, in the BNS mass range, at least in the case of zero spins, both the effectualness and faithfulness for any pair of PN waveforms and for any PN approximant with the EOB model tend to be above 0.99.\footnote{An exception is the so-called TaylorEt waveform, which is considered pathological for this reason.} For example, in the case of Advanced LIGO and for $(m_1, m_2) = (1.42,1.38)\,M_\odot$, the faithfulness of TaylorF2 against TaylorT4 is 0.999, and for TaylorF2 against EOB it is 0.996. 

The strong agreement between the various waveform approximants in the BNS mass range suggests that, at least for such systems, it is safe to adopt TaylorF2, the computationally least expensive waveform model, for the trial waveforms used in TIGER. However, since the pipeline is specifically meant to find small anomalies in the signals, we need to make sure that even small differences between waveform approximants are not mistaken for violations of GR.

\begin{figure}[!h]
\includegraphics[width=\columnwidth]{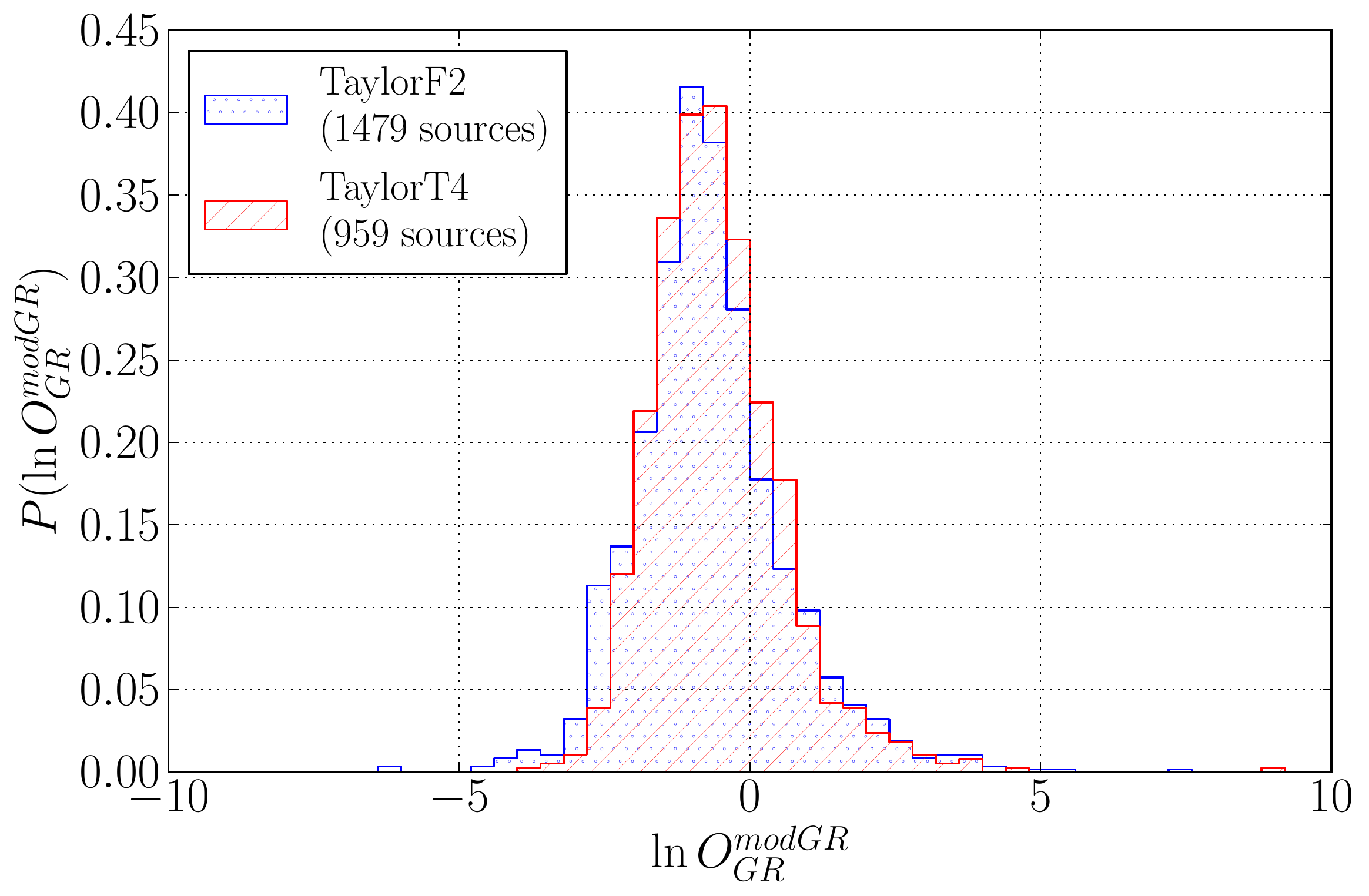}
\caption{Single-source background distributions for TaylorF2 injections (blue, dotted) and TaylorT4 injections (red, dashed), both analyzed with TaylorF2 waveforms cut off at 400 Hz.}
\label{fig:approximants}
\end{figure}

In Fig.~\ref{fig:approximants}, we compare single-source background distributions for the case where the GR signals are TaylorT4 waveforms and the case where they are TaylorF2 waveforms; but, in both cases, the analysis of the data is done with TaylorF2. Once again the difference between the two distributions can be quantified by using the KS statistic, which in this case comes out to be $D_{N,N'}^{\rm TF2, TT4} = 0.07$. 

Due to computational cost, we decided not to repeat the calculation with TaylorT1, TaylorT2, TaylorT3, or EOB injections. However, the results of Fig.~\ref{fig:approximants}, together with the waveform comparisons of \cite{Buonanno2009}, are sufficient to conclude that TIGER will not mistake differences in waveform models for violations of GR. 

\subsection{Effect of post-Newtonian order}

In \cite{Buonanno2009}, waveform approximants were considered up to 3.5PN in phase, which is the highest post-Newtonian order currently available. To this order, post-Newtonian waveforms and EOB-based inspiral-merger-ringdown waveforms that were tuned using numerical relativity simulations agree extremely well in the BNS mass regime. However, taking numerical relativity results to be the benchmark for how realistic a waveform model is, we note that large-scale numerical simulations of spacetimes containing coalescing binaries still only give information about the last few tens of cycles \cite{Ajith2012}, whereas a typical BNS waveform is thousands of cycles long. Thus, it could be that adequate modeling of the signals by post-Newtonian waveforms will require going to still higher PN order in the phase.\footnote{It is also possible that the contrary is true, since it is not known whether the post-Newtonian expansion converges \cite{Simone1997}.} 

\begin{figure}[!h]
\includegraphics[width=\columnwidth]{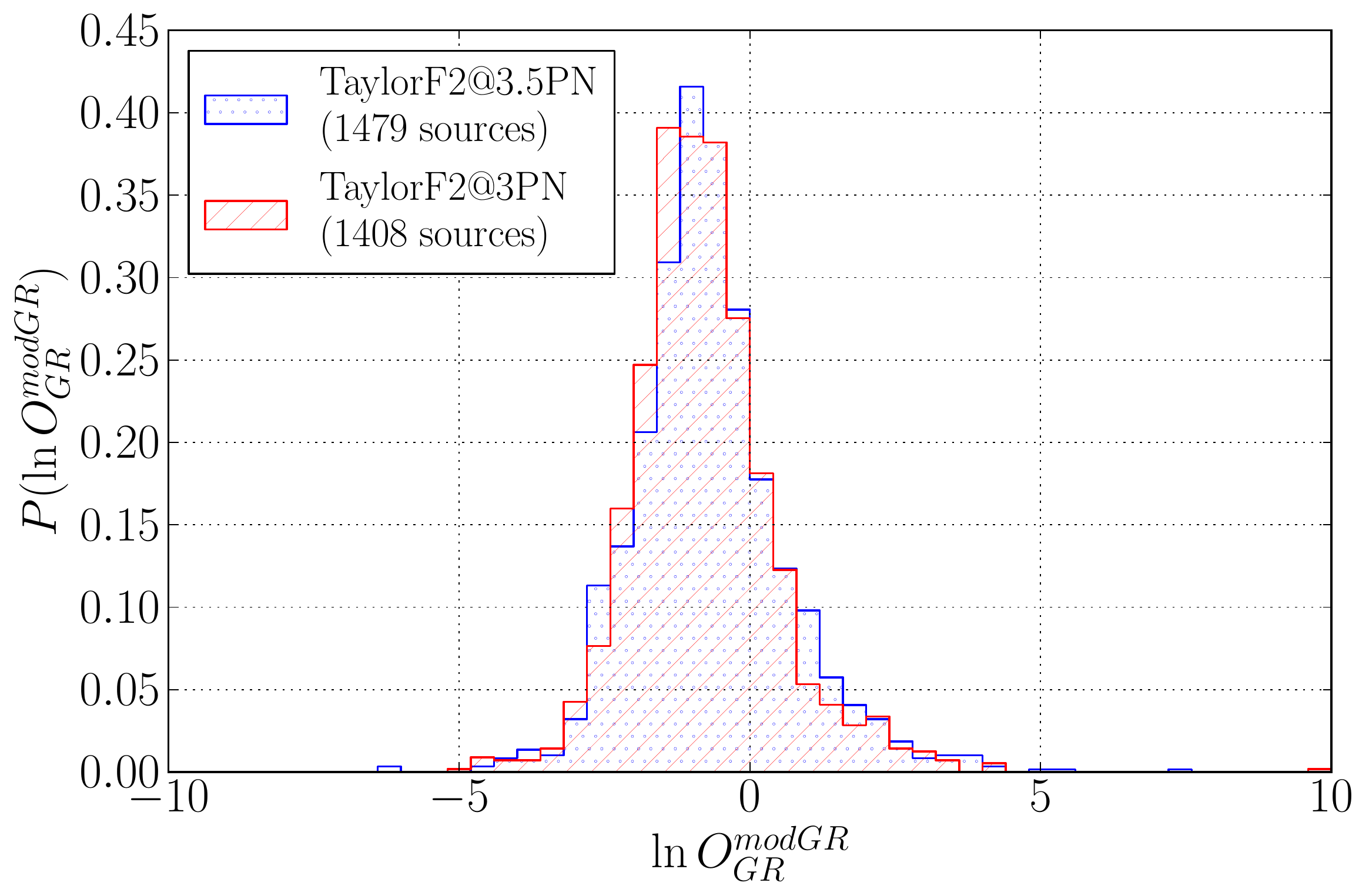}
\caption{Single-source background distributions for TaylorF2 injections to 3.5PN, where in one case the recovery waveform is also TaylorF2 to 3.5PN  (blue, dotted) and in the other case, TaylorF2 to 3PN (red, dashed), both cut off at 400 Hz.}
\label{fig:PNorder}
\end{figure}

In Fig.~\ref{fig:PNorder}, we probe the effect on the background of differences in post-Newtonian order between signal and recovery waveforms, for TaylorF2. In one case, both are taken to 3.5PN order, while in the other case the signal is 3.5PN whereas the recovery waveform only goes to 3PN. We see that the distributions barely differ; the KS statistic is $D_{N,N'}^{\rm 3PN, 3.5PN} = 0.05$. Needless to say, this does not prove that missing post-Newtonian orders beyond 3.5PN will be unproblematic, but it does lend further confidence to the soundness of our approach. Note also that one can expect high PN contributions to manifest themselves at high frequencies, and our recovery waveforms are cut off at 400 Hz.

\subsection{Instrumental calibration errors}

Imperfect calibration of the instruments can cause one to draw incorrect conclusions about detected signals. Calibration errors affect the instruments' transfer functions $R(f)$, which relate external length changes $\Delta L_{\rm ext}$ in the interferometer arms to the detector outputs $e(f)$:
\be
\Delta L_{\rm ext}(f) = R(f)\,e(f).
\ee
$R(f)$ is a complex function, which can be written in polar form as
\be
R(f) = \left[1 + \frac{\delta A}{A}(f)\right]\,e^{i\delta\phi(f)}\,R_e(f),
\ee
where $(\delta A/A)(f)$ and $\delta\phi(f)$ are frequency dependent calibration errors in amplitude and phase, respectively, and $R_e(f)$ is the transfer function in the absence of errors. The frequency domain data stream is given by 
\be
\tilde{d}(f) = \frac{\Delta L_{\rm ext}(f)}{L},
\ee
where $L$ is the interferometer arm length in the absence of disturbances. Calibration errors affect both the data stream $\tilde{d}$ and the power spectral density of the noise $S_n(f)$, but \emph{not} the model waveforms corresponding to the hypotheses $H_{i_1 i_2 \ldots i_k}$  and $\hyp_{\rm GR}$, which is how parameter estimation and model selection get affected by them.

In \cite{Vitale2012}, the calibration errors were modeled based on the errors measured in the initial LIGO and Virgo instruments, and their effect on Bayesian parameter estimation and model selection for advanced detectors was assessed. It was found that even with amplitude errors of $\delta A/A \sim 10\%$ and phase errors $\delta\phi \sim 3$ degrees in each instrument, for 90\% of sources the systematics induced will be less than 20\% of the statistical uncertainties in parameter estimation. Similarly, model selection is not much affected by calibration errors. 

\begin{figure}[!h]
\includegraphics[width=\columnwidth]{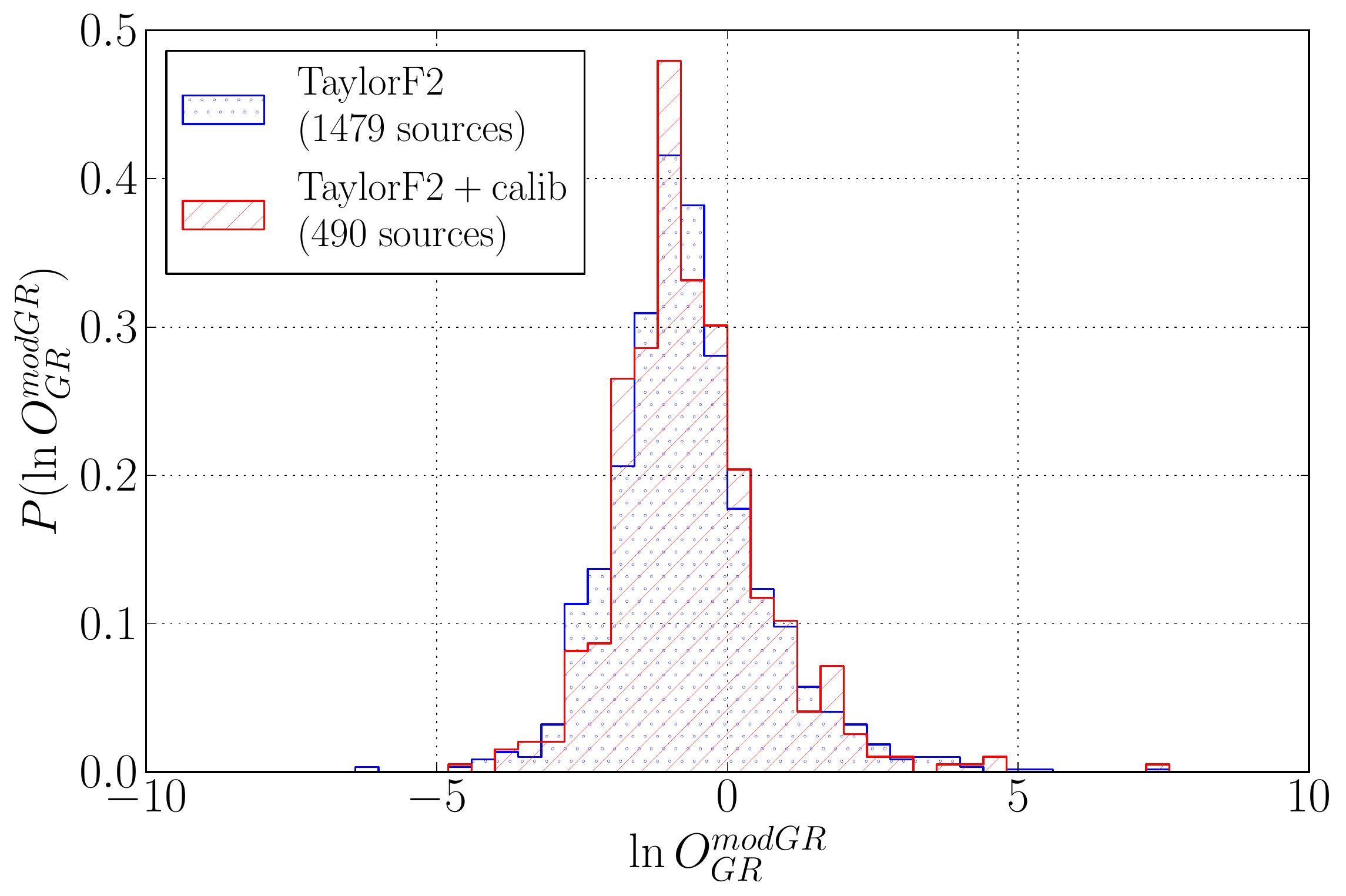}
\caption{Single-source background distributions for TaylorF2 injections without calibration errors (blue, dotted) and with frequency dependent amplitude and phase errors modeled as in \cite{Vitale2012} (red, dashed), both analyzed with TaylorF2 waveforms cut off at 400 Hz.}
\label{fig:calibration}
\end{figure}

Fig.~\ref{fig:calibration} shows the effect of calibration errors, modeled exactly as in \cite{Vitale2012}, on the log odds ratio background distribution. As expected, the effect is minor (with $D_{N,N'}^{\rm cal, nocal} = 0.04$), and calibration errors will not affect the performance of TIGER.

\subsection{Number of testing parameters}
\label{subsec:testingparameters}

\begin{figure}[!h]
\includegraphics[width=\columnwidth]{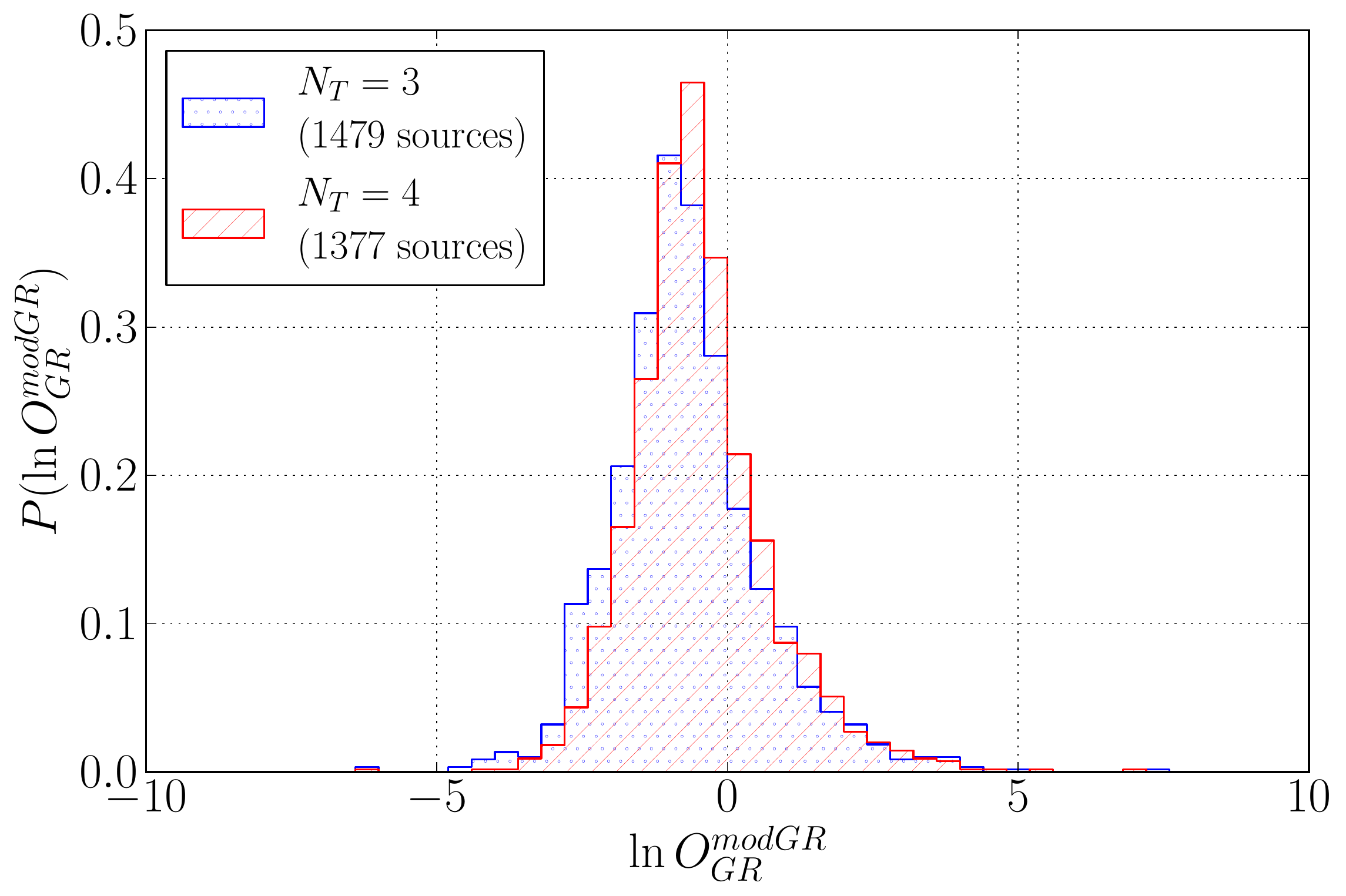}
\caption{Single-source background distributions for TaylorF2 injections with TaylorF2 recovery, in one case with three testing parameters (blue, dotted) and in the other with four (red, dashed).}
\label{fig:testingparameters}
\end{figure}

TIGER allows one to circumvent the usual problem in Bayesian analysis when the number of extra parameters in the model is too large: The total number of testing parameters, $N_T$, can in principle be arbitrarily large without risk of being penalized by the high dimensionality of the parameter space should the number of extra parameters in the signal be smaller than $N_T$. One aspect of this was already illustrated in \cite{Li2012a}, where it was shown that if the GR violation is limited to \emph{e.g.}~the 1.5PN phase coefficient, hypotheses with too many free parameters tend to be disfavored even if they include $\psi_3$. However, what also needs to be checked explicitly is how sensitive the background is to the number of testing parameters: Should it be the case that it widens dramatically as $N_T$ is increased because features in the noise can more easily be accommodated by waveforms with more free parameters, then the advantage disappears. In Fig.~\ref{fig:testingparameters}, we compare backgrounds for $N_T = 3$ and $N_T = 4$, and the difference 
turns out to be small; in terms of a KS statistic, $D^{3,4}_{N,N'} = 0.11$.

Together with the results of \cite{Li2012a}, this indicates that one should use as many testing parameters as possible. However, in practice there will be computational constraints due to the exponential growth of the number of sub-hypotheses with the total number of testing parameters; indeed, for $N_T$ testing parameters, $2^{N_T} - 1$ sub-hypotheses $H_{i_1 \ldots i_k}$ need to be compared with $\hyp_{\rm GR}$. The results of \cite{Li2012a,Li2012b} suggest that in the case of BNS, the sensitivity of TIGER to GR violations occurring above 2PN order in phase will be limited. In the examples below, we use three testing parameters, $\{ \psi_1, \psi_2, \psi_3 \}$. 

\subsection{Neutron star spins}

The observed pulsar spin periods and assumptions about neutron star spindown rates lead to periods at birth in the range 10--140 ms \cite{Lorimer2008}, corresponding to dimensionless spins $J/m^2 \lesssim 0.04$, and the fastest known pulsar in a BNS has a spin $J/m^2 \sim 0.02$. Thus, neutron star spins in BNS systems are generally expected to be small. Nevertheless, we need to quantify their effect on the background distribution and hence the detectability of GR violations. 

In the phase, spin-orbit effects first appear at 1.5PN order, and spin-spin effects at 2PN. The amplitude is also affected, primarily because of spin-induced precession of the orbital plane, which causes the inclination angle to change so that sometimes a system might be close to being face-on whereas at other times it will be closer to being edge-on, causing amplitude modulation. 

To describe the orbital motion with inclusion of spins, one again uses the Kepler and flux-energy balance equations, Eq.~(\ref{Kepler})--(\ref{energyflux}), with $E(v)$ and $\mathcal{F}(v)$ modified to take spin-orbit and spin-spin effects into account, and these are supplemented by a set of differential equations for the time evolution of the individual spins $\vec{S}_1$ and $\vec{S}_2$, and of the unit normal in the direction of orbital angular momentum, $\hat{L}$. For the purposes of this paper, spin effects were included to 2.5PN \cite{Faye2006}, although by now spin-orbit effects in the flux are known to 3.5PN \cite{Bohe2013}. In the case of spins that are (anti-)aligned with each other and the orbital angular momentum, so that there is no precession, it is not difficult to arrive at a closed expression for phase as a function of frequency in the stationary phase approximation \cite{Arun2009}. 

To assess the effect of spins, we constructed a background where the injected signals were TaylorT4 waveforms with precessing spins included in the dynamics, as described above. (Results for injections with (anti-)aligned spinning TaylorF2 waveforms were already reported in \cite{MGproceedings}.)  The spin orientations were picked from a uniform distribution on the sphere, and their magnitudes followed a Gaussian distribution centered on zero and with $\sigma = 0.05$. The recovery waveforms were again TaylorF2, but this time allowing for spins that are aligned or anti-aligned with orbital angular momentum. We need to pick a prior distribution for the spin magnitudes in the recovery waveform. In the present setting, the most natural choice is again a Gaussian centered on zero and having a width of 0.05. Indeed, letting spins in the recovery waveform vary within a wide range could lead us to miss GR violations occuring from 1.5PN order onward, since such deviations could be accomodated by adjusting the spins. 

We explicitly note that the smallness of neutron star spins is an astrophysical assumption that enters the background calculation; see Sec.~\ref{sec:discussion} below for a discussion. However, given general astrophysical considerations as well as currently observed binary neutron star systems \cite{Lorimer2008}, most likely our choice of spin distributions in injections and recovery waveforms leads to a background that is rather conservative. 

\begin{figure}[!h]
\includegraphics[width=\columnwidth]{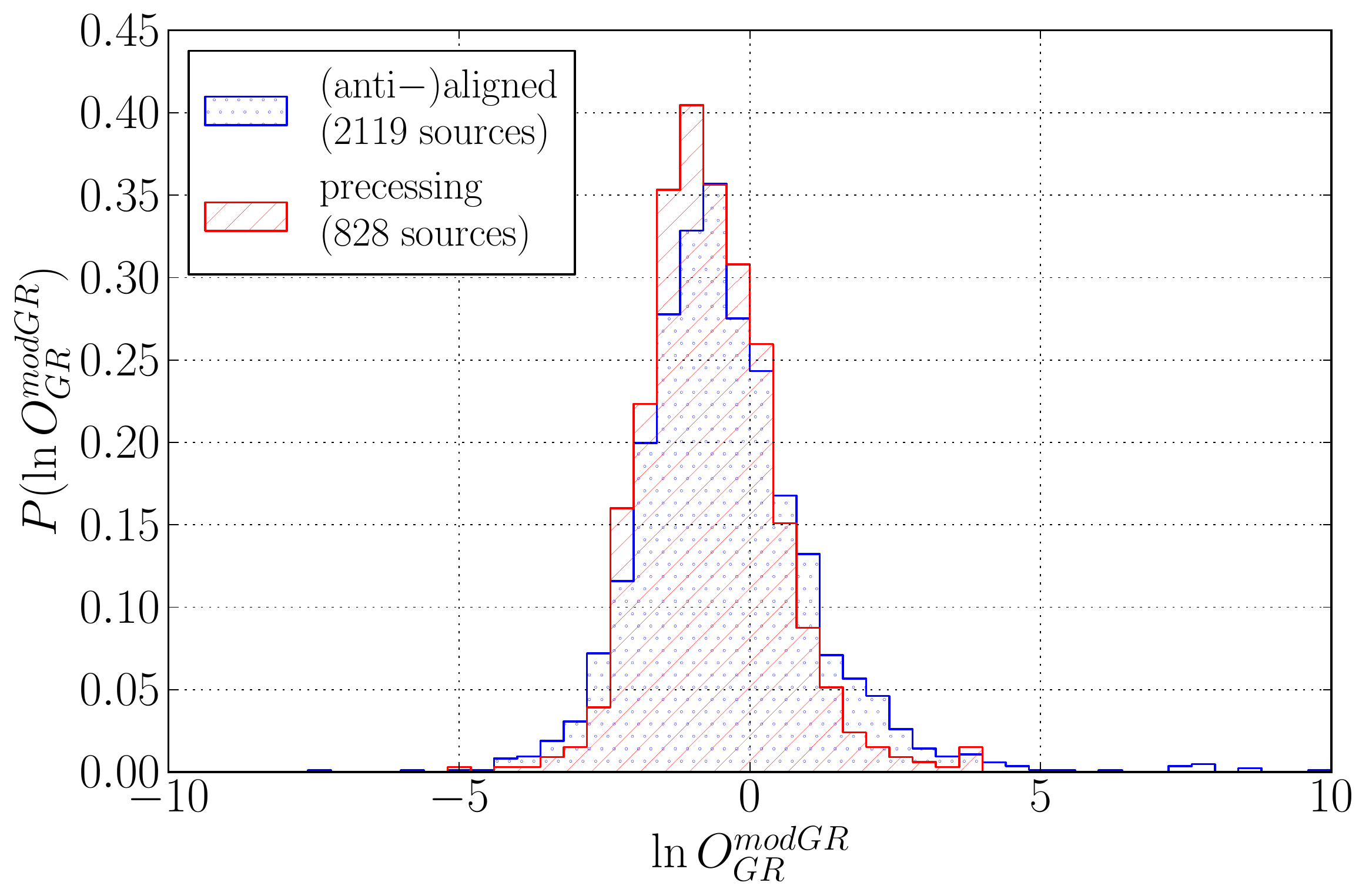}
\caption{Single-source background distributions for TaylorF2 injections with (anti-)aligned spins (blue, dotted) and TaylorT4 injections with precessing spins (red, dashed). In both cases the recovery is with TaylorF2 waveforms cut off at 400 Hz.}
\label{fig:spins}
\end{figure}

Since the injections have precessing spins while in the recovery we only allow for (anti-)aligned spins, the recovery waveform model will not perfectly capture the signal even for BNS. Nevertheless, the effect on the background distribution is minor, as shown in Fig.~\ref{fig:spins}; one has $D^{\rm align, prec}_{N,N'} = 0.08$. Clearly, given the relative smallness of the spins, allowing for (anti-)aligned spins in TaylorF2 is sufficient for this waveform model to capture the spin effects in the signal, at least to the extent that the background is not significantly affected. 

\subsection{Combined effect of differences between waveform approximants, tidal deformation, calibration errors, and spins}

We now put everything together and compute a background distribution where the recovery waveform is TaylorF2 with (anti-)aligned spins, cut off at 400 Hz, but the injections are TaylorT4 with precessing spins and tidal effects at 0PN and 1PN, and calibration errors are also included. In the case of TaylorT4, the phase is only computed numerically, and tidal effects must be added in the equation for $dv/dt(v)$:
\be
\frac{dv}{dt}(v) = \mathcal{G}_{\rm PP}(v) + \mathcal{G}_{\rm tidal}(v),
\ee
where to 1PN order \cite{Vines2011} 
\ba
&&  \mathcal{G}_{\rm tidal}(v) \nn\\
&&= \frac{16\chi_1\lambda_2}{5 M^6}\,\left[ 12(1+11\chi_1)\,v^{19}  \right. \nn\\
&&+ \left.  \left( \frac{4421}{28} - \frac{12263}{28} \,\chi_2 + \frac{1893}{2}\,\chi_2^2 - 661\,\chi_2^3 \right)\, v^{21}\right] \nn\\
&&+\,\, (1 \leftrightarrow 2). 
\ea
For the expression of the point particle contribution $\mathcal{G}_{\rm PP}(v)$ to 3.5PN, with spins included up to 2.5PN, we refer to \cite{Faye2006}.

\begin{figure}[!h]
\includegraphics[width=\columnwidth]{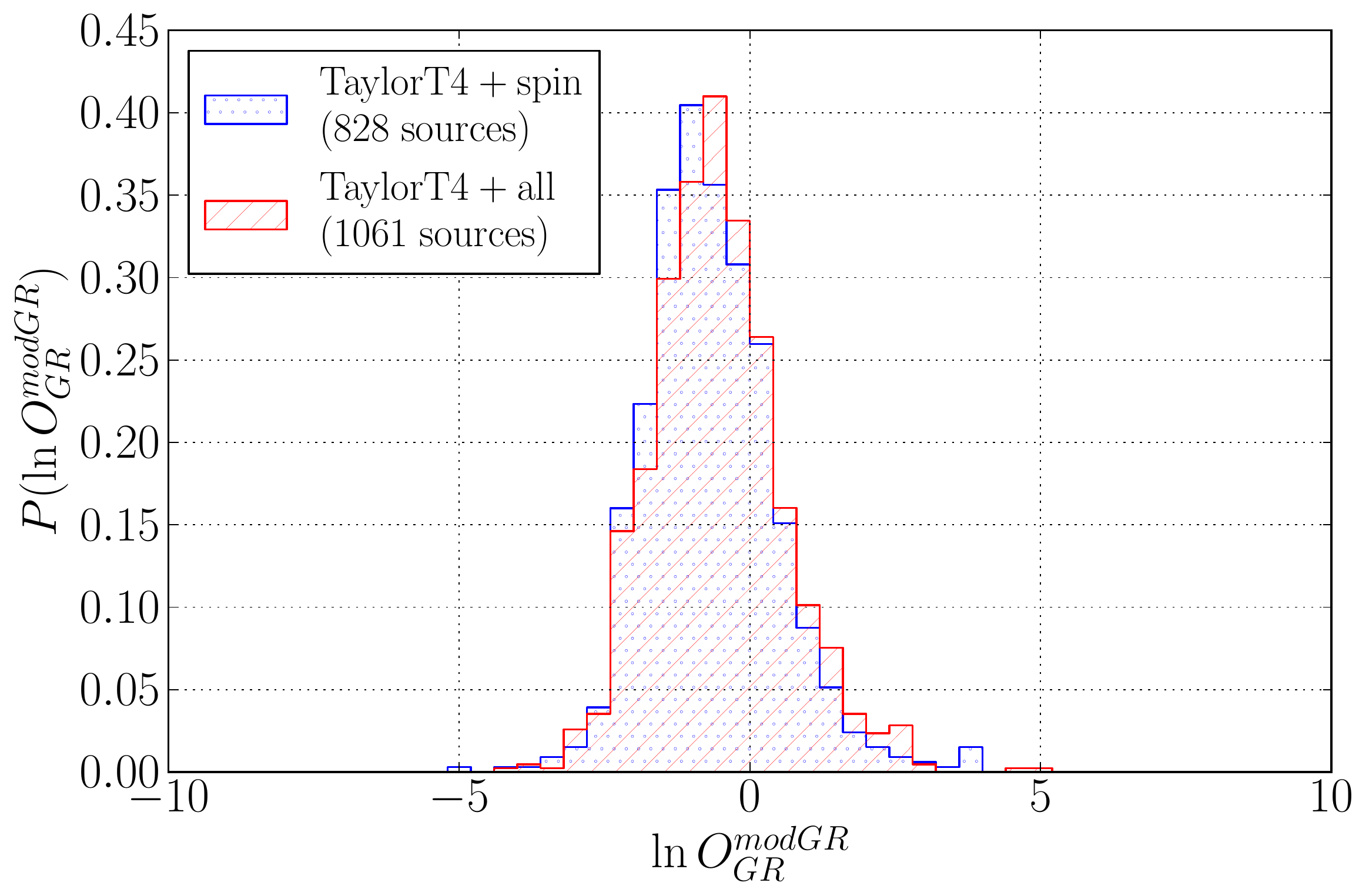}
\caption{Single-source background distributions for TaylorT4 injections with precessing spins (blue, dotted) and TaylorT4 injections with precessing spins, tidal effects, and calibration errors (red, dashed). In both cases, the recovery is with (anti-)aligned spinning TaylorF2 cut off at 400 Hz.}
\label{fig:ultimate}
\end{figure}

For the case of single sources, the effect on the background of a combination of precessing spins, tidal effects, and calibration errors is shown in Fig.~\ref{fig:ultimate}. In terms of a KS statistic, the difference between backgrounds is $D_{N,N'}^{\rm spins, all} = 0.07$.

\begin{figure}[!h]
\includegraphics[width=\columnwidth]{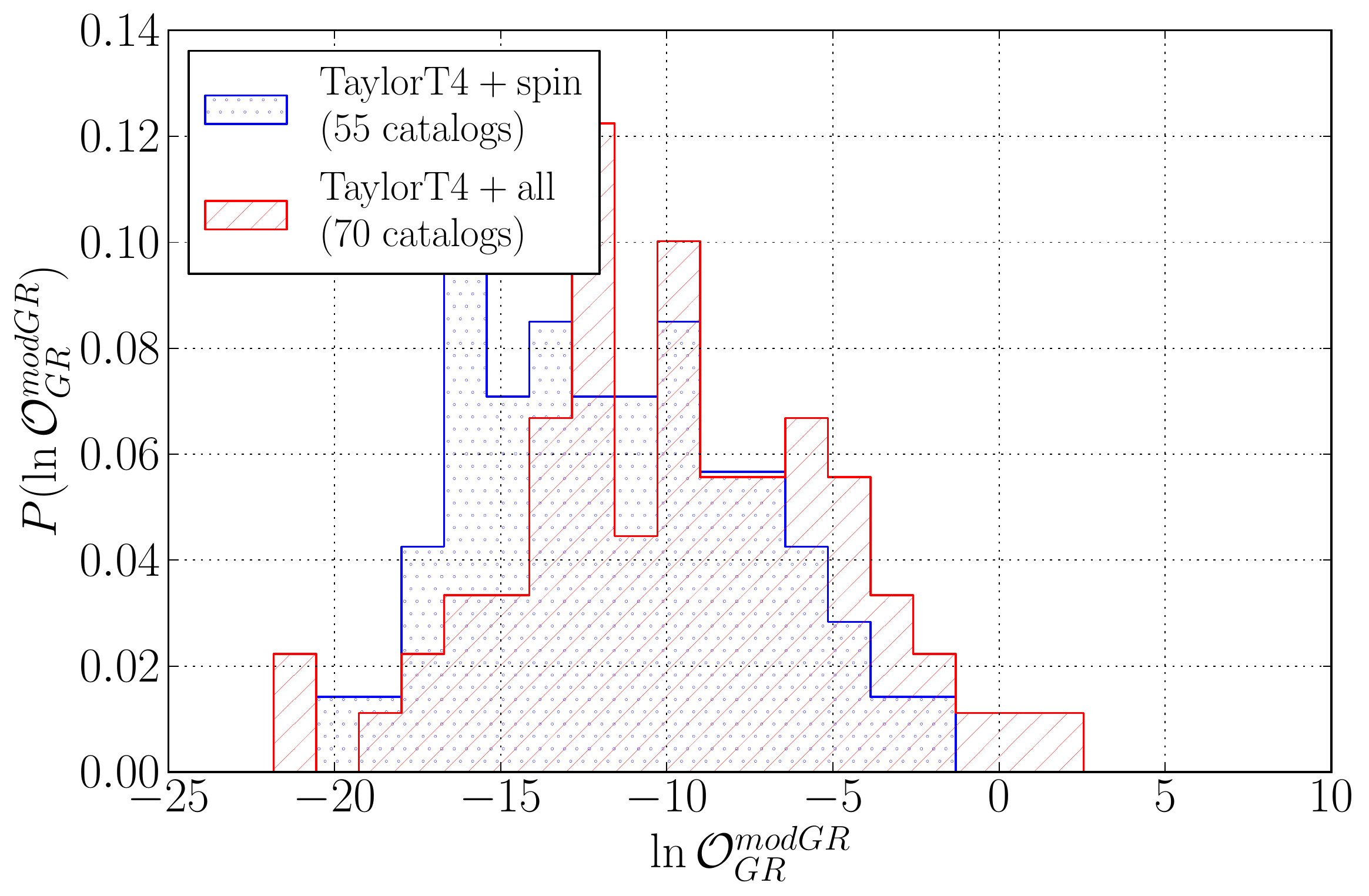}
\caption{The same comparison as in Fig.~\ref{fig:ultimate}, but now for \emph{catalogs} of 15 sources each. Note how GR is typically much more favored when information from multiple GR sources is combined.}
\label{fig:ultimatecatalogs}
\end{figure}

For reasons of computational expense, so far we have only shown differences between backgrounds for \emph{single sources}, which is appropriate for the case where there is only one detection. If there are $\mathcal{N}$ detections that can be clearly identified as BNS events according to the criterion $\mathcal{M} < 1.3\,M_\odot$, then one will want to construct a background distribution for \emph{catalogs} of $\mathcal{N}$ sources each. We computed backgrounds using the injection sets of Fig.~\ref{fig:ultimate}, but now randomly combining injections into catalogs of 15 sources each. The results are shown in Fig.~\ref{fig:ultimatecatalogs}. When information from multiple GR sources is combined, one expects $\hyp_{\rm GR}$ to be much more favored over $\hyp_{\rm modGR}$, and this is what we see: in both cases, the distribution of $\ln\mathcal{O}^{\rm modGR}_{\rm GR}$ stretches to much more negative values. However, when making comparisons of different physical set-ups, combining information from multiple sources can make the differences show up much more clearly than in the case of single sources. For the purposes of this paper, a much smaller number of simulations were performed than one would in reality; one has ${}^{\rm (cat)}D_{N,N'}^{\rm spins, all} = 0.24$, but this will in large part be due to small number statistics. Reassuringly, even for catalogs of sources, the two background distributions are rather similar, with both favoring strongly negative values of log odds. 

Finally, we want to show at least one example of how well violations of GR might be detectable in the presence of strong tidal effects, instrumental calibration errors, and precessing spins. Recalling that the 1.5PN contribution to the orbital motion is where, according to GR, the dynamical self-interaction of spacetime first becomes visible \cite{Blanchet1994,Blanchet1995}, we consider a (heuristic) violation of GR at that order, taking the form of a $-10\%$ shift in the relevant coefficient in the expansion of $dv/dt(v)$:
\ba
\frac{dv}{dt}(v) &=& \mathcal{G}_{\rm PP}(v) + \mathcal{G}_{\rm tidal}(v) \nn\\ 
&& + \,\delta\xi_3\,\alpha_3(m_1, m_2, \vec{S}_1, \vec{S}_2)\,v^{12},
\ea
where we note that the leading-order contribution to $dv/dt$ goes like $v^9$; $\alpha_3(m_1, m_2, \vec{S}_1, \vec{S}_2)$ is the 1.5PN coefficient predicted by GR, and $\delta\xi_3 = -0.1$. 

\begin{figure}[!h]
\includegraphics[width=\columnwidth]{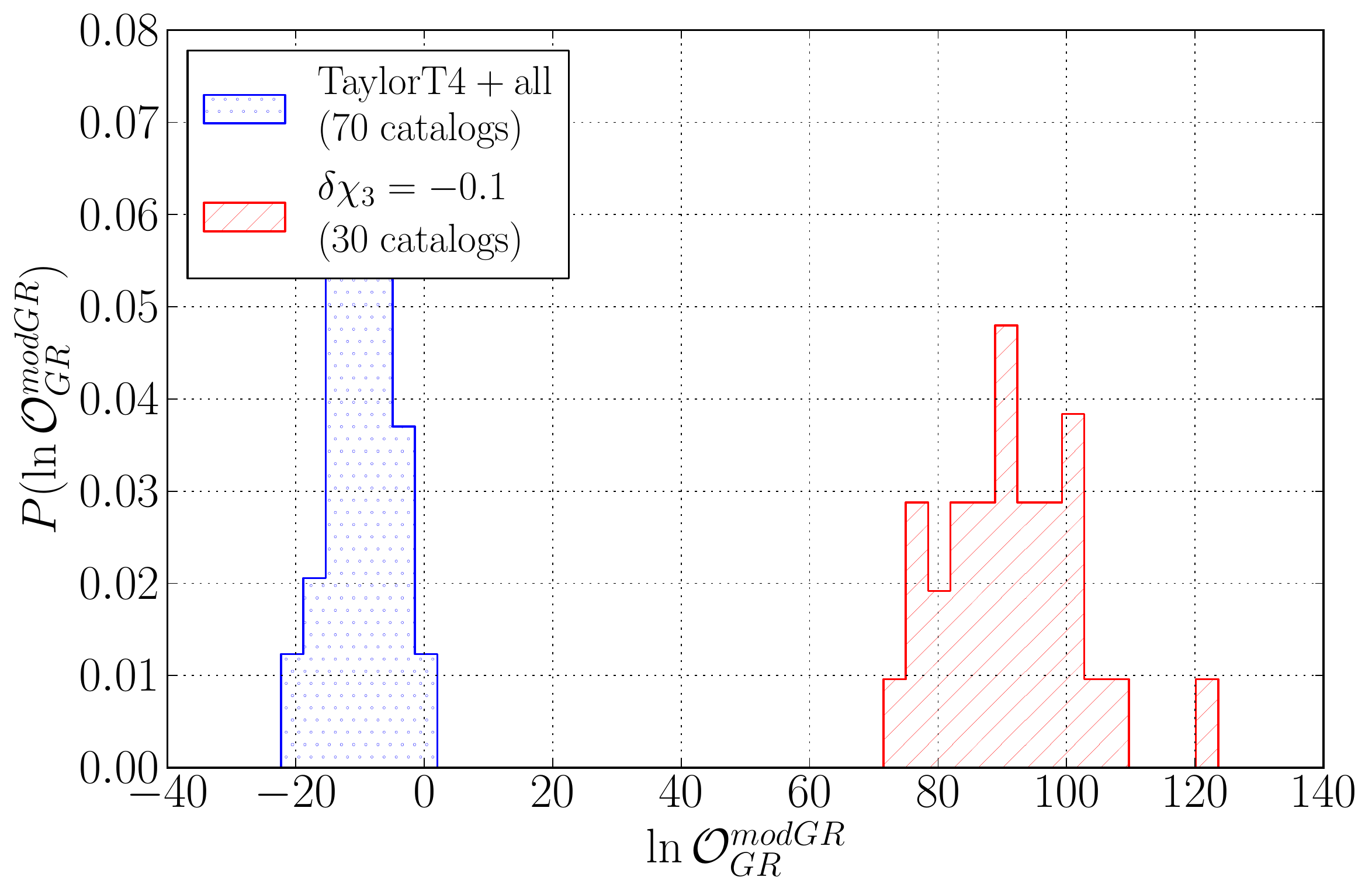}
\caption{Log odds ratio distributions for catalogs of 15 sources each. The blue, dotted histogram is the GR background for TaylorT4 signals with precessing spins, neutron star tidal deformation, and instrumental calibration errors. The red, dashed one is a foreground distribution for signals with the same effects present, and with a GR violation that takes the form of a constant $-10\%$ shift at 1.5PN, as explained in the main text. In both cases, the recovery is with (anti-)aligned spinning TaylorF2 waveforms cut off at 400 Hz.}
\label{fig:ultimateall}
\end{figure}

In Fig.~\ref{fig:ultimateall}, we show background as well as foreground log odds ratio distributions, for catalogs of 15 sources each, where in both cases the injections include neutron star tidal deformation, instrumental calibration errors, and precessing spins. As before, the recovery is with TaylorF2 waveforms that allow for (anti-)aligned spins, cut off at a frequency of 400 Hz. We see that the separation between the distributions is complete: almost regardless of false alarm probability, with 15 BNS detections the efficiency in finding the given GR violation is essentially 100\%.

\section{Conclusions and future directions}
\label{sec:discussion}

We have developed TIGER, a data analysis pipeline to perform model-independent tests of general relativity in the strong-field regime, using detections of compact binary coalescence events with second-generation gravitational wave detectors. The basic idea is to compare the GR hypothesis $\hyp_{\rm GR}$ with the hypothesis $\hyp_{\rm modGR}$ that one or more coefficients in the post-Newtonian expression for the phase do not depend on component masses and spins in the way GR predicts. Though the latter hypothesis has no waveform model associated with it, it can be written as the logical union of mutually exclusive sub-hypotheses, in each of which a fixed number of phase coefficients are free parameters on top of component masses, spins, sky position, orientation, and distance, while the others depend on masses and spins in the way GR predicts. In present form, the pipeline can in principle already be applied to binary neutron star events, for which waveform models that are reliable and can be generated sufficiently fast on a computer are available.

We performed a range of numerical experiments to check the robustness of TIGER against fundamental, astrophysical, and instrumental unknowns. In the BNS mass regime, the differences between the available waveform approximants are very small, making it unlikely that imperfect modeling of the signal will cause us to suspect a violation of GR. The fact that waveforms are only known up to a finite post-Newtonian order should also not be cause for concern. In the final stages of inspiral, finite size effects are important and the neutron stars will deform each other in an essentially unknown way; however, if the recovery waveforms are cut off at 400 Hz then the unknown tidal effects will not be mistaken for violations of GR, but the performance of TIGER remains unaffected. Instrumental calibration errors of expected size will not be problematic. Finally, if, as generally expected, the spins of neutron stars in binaries are small, then they can easily be dealt with. 

In present form, TIGER relies on two important astrophysical assumptions. One is that NSBH and BBH coalescences have chirp masses above a certain value, so that such events can be discarded, leaving only BNS. The other is the relative smallness of spins for BNS. In the future we will also want to work with BBH and NSBH events so that if an anomaly is discovered in BNS signals, we can confirm that it is of a fundamental rather than an astrophysical nature by using qualitatively different systems. Pan \emph{et al.}~appear to have arrived at a reliable semi-analytic waveform model for BBH and NSBH coalescence \cite{Pan2013}, and their approximant will be extremely useful as an injection waveform. However, it is too computationally expensive to be used for recovery. On the other hand, very recently Hannam \emph{et al.}~\cite{Hannam2013b} proposed a \emph{frequency domain} inspiral-merger-ringdown waveform which captures precessing spins, and which may already be useful for our purposes. An upgrade of the fast time domain ``PhenSpin" waveform of Sturani \emph{et al.} could also be an option for recovery \cite{PhenSpin1,PhenSpin2}. (Note that for the 
background calculation, it is important that the \emph{injected} waveform model be as close as possible to reality, but the requirements for the recovery waveform are less stringent.) To have some idea of what might conceivably be possible with BBH, we used the earlier BBH waveform approximant of \cite{Santamaria2010} with spins set to zero, for both injection and recovery, choosing component masses to be in the range $[5, 15]\,M_\odot$ and placing sources uniformly in co-moving volume with distances up to 1.25 Gpc. It was found that for catalogs of 20 sources each, a deviation in (the equivalent of) the 3PN phase coefficient $\psi_6$ of only 0.5\% could be picked up with essentially 100\% efficiency, using only $\{ \psi_1, \psi_2, \psi_3, \psi_4 \}$ as testing coefficients; see Fig.~6 of \cite{VanDenBroeck2013}. Here some caution is called for, considering that astrophysical black holes are likely to have large, non-aligned spins, but the result is encouraging. The possibility of reliably applying TIGER to BBH detections 
using a waveform model along the lines of Hannam \emph{et al.}~\cite{Hannam2013b} or Sturani \emph{et al.} \cite{PhenSpin1,PhenSpin2} will be a subject of intense investigation.


Demonstrating the robustness of TIGER, applied to BNS, against fundamental and astrophysical unknowns as well as instrumental calibration errors was a necessary first step in determining whether it will be viable as a data analysis pipeline. A crucial further check will be to assess the behavior of TIGER in real noise, which is not quite stationary or Gaussian. We are in the process of testing the pipeline using existing data taken by the initial LIGO and Virgo detectors, but ``recolored" so that the underlying power spectral densities are the ones predicted for the advanced interferometers, while retaining the non-stationarities in the noise. Results will be reported in a forthcoming publication.

\section*{Acknowledgements}

MA, WDP, TGFL, CVDB, and JV were supported by the research programme of the Foundation for Fundamental Research on Matter (FOM), which is partially supported by the Netherlands Organisation for Scientific Research (NWO). SV acknowledges the support of the National Science Foundation and the LIGO Laboratory. LIGO was constructed by the California Institute of Technology
and Massachusetts Institute of Technology with funding from the National Science Foundation and operates under cooperative agreement PHY-0757058. The authors would like to acknowledge the LIGO Data Grid clusters, without which the simulations could not have been performed. Specifically, these include the computing resources supported by National Science Foundation Grants PHY-0923409 and PHY-0600953 to UW-Milwaukee. Also, we thank the Albert Einstein Institute in Hannover, supported by the Max-Planck-Gesellschaft, for use of the Atlas high-performance computing cluster. It is a pleasure to thank E.~Berti, A.~ter Braack, A.~Buonanno, N.~Cornish, J.D.E.~Creighton, W.M.~Farr, B.R.~Iyer, C.K.~Mishra, C.~Pollice, B.S.~Sathyaprakash, R.~Sturani, and N.~Yunes for useful discussions.

%
%

\end{document}